\documentstyle[12pt]{article}
\newcommand{\sect}[1]{\section{#1}\setcounter{equation}{0}}

\newcommand{\eq}{\begin{equation}}
\newcommand{\eqe}{\end{equation}}
\newcommand{\dis}{\begin{displaymath}}
\newcommand{\dise}{\end{displaymath}}
\newcommand{\eqa}{\begin{eqnarray}}
\newcommand{\eqae}{\end{eqnarray}}
\newcommand{\e}[1]{\label{eq:#1}}
\newcommand{\ee}[1]{(\ref{eq:#1})}

\newcommand{\apm}{\alpha'}
\newcommand{\half}{\mbox{$\frac{1}{2}$}}
\newcommand{\roughly}[1]%
{\raise.3ex\hbox{$#1$\kern-.75em\lower1ex\hbox{$\sim$}}}

\begin{document}\bigskip
\hskip 3.7in\vbox{\baselineskip12pt
\hbox{NSF-ITP-97-097}\hbox{hep-th/9707170}}
\bigskip\bigskip\bigskip

\centerline{\large \bf  Self Gravitating Fundamental Strings}

\bigskip\bigskip

\centerline{\bf Gary T. Horowitz}
\medskip
\centerline{Physics Department}
\centerline{University of California}
\centerline{Santa Barbara, CA  93106}
\medskip
\centerline{email: gary@cosmic.physics.ucsb.edu}
\bigskip
\centerline{\bf Joseph Polchinski}
\medskip
\centerline{Institute for Theoretical Physics}
\centerline{University of California}
\centerline{Santa Barbara, CA. 93106}
\medskip
\centerline{e-mail: joep@itp.ucsb.edu}
\bigskip\bigskip

\begin{abstract}
\baselineskip=16pt
We study the configuration 
of a typical highly excited string as one slowly
increases the string coupling. The dominant interactions are the long
range dilaton and gravitational attraction. 
In four spacetime dimensions, the string
slowly contracts from its initial (large) size until it approaches 
the string scale where
it forms a black hole. In higher dimensions, the string stays large
until the coupling reaches a critical value, and then it rapidly collapses to
a black hole. The implications for the recently proposed correspondence
principle are discussed.

\end{abstract}
\newpage
\baselineskip=18pt

\sect{Introduction}

We have recently formulated a {\it correspondence
principle} which relates black holes and strings~\cite{horpol}.
Developing ideas in~\cite{suss1}, it was proposed that as one adiabatically
decreases the string coupling $g$, a black hole makes a transition to a
state of weakly coupled strings (and D-branes) with the same mass, charge and
angular momentum as the black hole. For most black holes in string
theory, namely those without magnetic Neveu-Schwarz (NS) charges, the ratio 
of the horizon size to the fundamental string length $\apm^{1/2}$ decreases 
as one decreases $g$. The transition
occurs when this ratio is of order one.
Before this point,
the black hole is well defined as a solution of the low energy supergravity
theory; beyond this point, $\alpha'$ corrections become large and the metric
near the horizon is no longer well defined. 

By relating a black hole to a set of weakly coupled states,  
the correspondence principle provides a statistical description of black hole
entropy. We have verified that the number of such weakly coupled states 
agrees with the Bekenstein-Hawking entropy in a wide variety of
examples involving Ramond and electric NS charges
in various dimensions.
In contrast to the precise counting of states for extremal
and near-extremal black holes~\cite{stromvafa,entreviews}, this method
does not in general determine the
numerical coefficient in the entropy since that would depend on the precise
coupling at which the transition occurs.  However, it applies to a much
wider class of black holes and reproduces the correct
functional dependence on the mass and charges. Other aspects of this
correspondence principle have been investigated recently in~\cite{newcp}.

In the present paper we would like to develop this further by considering
the reverse process; we start with weakly coupled
matter and increase the  string coupling.  We focus on the simplest case of a
single highly excited string (in various dimensions).  
The physics of free highly excited strings has been much discussed in
connection with the Hagedorn transition.  One of our results is to include
string interactions in the behavior of the string.  One might hope that
this would ultimately lead to a better understanding of the Hagedorn
transition, but it does not apply directly because we consider only a single
isolated string.

Consider a string state at level $N \gg 1$, with mass $M^2 = 4N/\apm$.
As we increase the string coupling, the effective Schwarzschild
radius $GM$ increases. It becomes of order the string
scale when $g^2 N^{1/2} \sim 1$. This defines a critical coupling
when the transition  to a black hole can first occur:
\eq
g_{c} \sim N^{-1/4}\ . \e{critg}
\eqe
Notice that $g_c \ll 1$ and is independent of the number of (noncompact) spatial
dimensions. Conversely, if one starts with a Schwarzschild black hole
and decreases the coupling,  the horizon size will be
of order the string scale when the coupling is given by \ee{critg},
where $N\sim M_{bh}^2\apm$. So this is the coupling at which the correspondence
principle predicts the transition between black holes and strings.

However, a string at level $N$ will form a black hole when
$g\sim g_c$ only if it
is confined to about the string scale.
At zero coupling, the typical size of the string is
much larger: $\ell = N^{1/4} \apm^{1/2}$.
This follows from the random walk picture of the
excited string~\cite{rand}, where it takes $N^{1/2}$ steps each of
length $\apm^{1/2}$. The key question is what happens to this size as we
increase $g$. Intuitively, one would expect gravitational and dilaton
forces to cause the string state to shrink, but it is not obvious that
it will shrink all the way to $ \ell \sim \apm^{1/2}$ by the critical coupling
\ee{critg}. Surprisingly, it turns out that the behavior of the string 
as one increases $g$
depends crucially on the number $d$ of (noncompact) spatial dimensions. 
We will see that interactions remain negligible until the coupling is
of order
\eq
g_o \sim N^{(d-6)/8}\ .
\eqe
As expected, the 
interactions always become important before the Schwarzschild radius reaches
the original random walk size, which occurs at a coupling $g \sim N^{(d-4)/8}$.

Four spacetime dimensions ($d=3$) yields perhaps the simplest behavior. 
In this case, $g_o < g_c$
so as one increases the coupling, the interactions first become
important and cause $\ell$ to decrease. In fact, we find that
\eq
\ell\sim {\apm^{1/2}\over g^2 N^{1/2}}
\eqe
so that as $g$ increases from $g_o$ to $g_c$, $\ell$ smoothly
contracts from the random walk size to the string scale. 
For $d=4$, $g_o = g_c$ so as one increases the coupling the string remains
large until $g$ approaches this value, and then it 
collapses to form a black hole. For $d=5$, $g_o > g_c$.  This leads to a
form of hysteresis. If we start with a  typical highly excited string
state and increase $g$, it will remain large until $g\sim g_o$,
at which point it
collapses into a black hole. If we now decrease the coupling, the black hole
remains a good description until $g\sim g_c$ at which point it turns into
an excited string state\footnote{For $g_c < g < g_o$, there is a
very small probability that the large string will undergo a fluctuation
to a small volume and become a black hole. There is also a very small
probability that the black hole will Hawking radiate a large string.
In addition, both the long string and black hole slowly lose mass by
radiating light particles. Since we are ignoring these effects, 
our adiabatic change in $g$ should not be so slow that the long string and
black hole become unstable.}.
 For $d= 6$, typical
excited string states remain large until $g \sim 1$ when
other strong coupling effects are likely to become important. 
The cases $d >6 $ can probably be analyzed by passing to a weakly coupled
dual description.

We will derive the above results in the next two sections
using a thermal scalar
formalism~\cite{therm}, which has been applied previously to try to understand
the critical behavior near the Hagedorn transition.
(See \cite{loth} for
another approach to include string interactions.) However first we
discuss their implications for the correspondence principle.
At first sight, the fact that typical string states do   not evolve into
black holes at the critical coupling $g_c$ in $d>4$ seems to contradict
both the explanation of black hole entropy
and the assumed reversibility of the adiabatic change in $g$.
However this is not the case. 
The resolution, which was mentioned briefly in~\cite{horpol}, is that 
as one decreases $g$, a higher dimensional black hole
becomes a highly excited string but in an {\it atypical} state. 
It must
still be long, with a length of order $ N^{1/2}\apm^{1/2} \sim M_{bh} \apm$ 
since we would expect of order
half of its energy to be in the form of string tension.  But rather than a
random walk, it is constrained to lie in a volume roughly set by the string
scale.  This is plausible because the correspondence principle should still
hold if the black hole is placed in a box only slightly larger than its own
size, which near
$g_{c}$ is the string length.

Are there enough of these atypical states to account for the black hole
entropy? For random-walking
strings the log of the number of states should be the number of steps, $N^{1/2}$
times a numerical constant.  This is indeed the entropy of highly
excited strings.  But this should also hold for the random walk constrained to
lie in a small volume. 
Compare random walks on an infinite two-dimensional square
grid and a small grid, say two squares by two.  The first walk has 4 choices
at each step and an entropy $N^{1/2} \ln 4$.   The second has 4, 3, or 2 at each
step, depending on whether the path is in the interior, at an edge, or at a
corner, and so the entropy will be $N^{1/2} \ln O(3)$.
The numerical coefficient is
outside the accuracy of the correspondence principle in any event.\footnote
{Note however that this coefficient appears in the
exponent in the number of states,
so the actual number of constrained random walks is much less than unconstrained
walks.} The net result is that the black hole evolves to a typical
excited string state only in three and four spatial dimensions. But
in all dimensions,
{\it the reversible adiabatic transition at $g\sim g_{c}$ is between black holes
and long but compact strings}.  

The string states associated with large random walks should also contribute
to the Bekenstein-Hawking
entropy when they form a black hole at larger values of the coupling,
but this is a small correction. 
For a given level $N$, 
a black hole which forms at $g=g_o$ has a larger mass in Planck units
than a black hole which forms at\footnote{They, of course, have the same
mass in string units, but it is the black hole mass in Planck units which
remains constant as $g$ is varied.} $g=g_c<g_o$. The dominant contribution
to the entropy of this larger black hole comes from compact strings
with $N'>N$ which make the transition when $g\sim g_c\sim {N'}^{-1/4}$. 
For example, in $d=5$, a string at level $N$ forms a black hole at $g\sim g_o
\sim N^{-1/8}$
with mass
\eq
M_{bh} \sim {N^{1/2} g_o^{1/2}\over l_p} \sim {N^{7/16}\over l_p}
\eqe
where $l_p$ is the Planck length.
If we now decrease the coupling to $g\sim g_c$, the black hole will form an
excited string with mass
\eq
{{N'}^{1/2}\over \apm^{1/2}} \sim {N^{7/16}\over (g_c \apm)^{1/2}}
\eqe
which implies $N' = N^{7/6}$.

In the next section we review the properties of highly excited free strings,
using the thermal scalar formalism. In section three we include interactions
by first considering the string in a fixed metric and dilaton background,
and then requiring that the background satisfy the equations of motion
with the typical excited string as source. The appendix includes some
details of the calculation of the stress energy tensor of the string.

\sect{Highly Excited Free Strings}

We are interested in the properties of a typical string state of mass $M \gg
\apm^{-1/2}$, given by the microcanonical ensemble.  However, it is easier to
calculate in the canonical ensemble, and so we will do this
and then solve for the mass in terms of the temperature.
Consider the one-string expectation value of some quantity $X$,
\eq
\langle X \rangle = Z^{-1} {\rm Tr}\,(X e^{-\beta H})\ ,
\quad Z = {\rm Tr}\,(e^{-\beta H})\ .
\eqe
As is well known, there is a limiting (Hagedorn) temperature beyond which the
traces diverge~\cite{hagedorn}.  This divergence is due to the exponential
rise in the density of states, $n(M) \sim e^{\beta_{\it H} M}$ where the inverse
Hagedorn temperature is of order the string scale:
$\beta_{\it H} \sim \apm^{1/2}$.
The critical behavior as
$\beta \to \beta_{\it H}$ is governed by strings with $M \gg \apm^{-1/2}$, and
so the properties of these strings can be extracted from the
critical behavior.

This critical behavior can be described by an effective field
theory of a single complex scalar field in one fewer spacetime
dimension~\cite{therm}.  This can be understood as follows.  The string
partition function can be calculated from a path integral in Euclidean
time with period
$\beta$.  Let us make a Euclidean rotation so that
instead we are considering the zero-temperature behavior with a spatial
dimension compactified.  The Hagedorn singularity then appears at a
critical compactification radius.  Such a singularity must arise from a
field becoming massless.  In this case it is a scalar of winding number
one which becomes tachyonic for $\beta < \beta_{\it
H}$,\footnote{This tachyon is present even in supersymmetric
string theories, because the thermal boundary conditions imply that 
spinors are anti-periodic, which breaks
supersymmetry.}
\eq
m^2(\beta) = \frac{\beta^2 - \beta_{\rm H}^2}{4\pi^2\apm^2}\ .
\eqe

The critical behavior of the free string partition function is thus
given by the thermal scalar path integral
\eq
{\bf Z} = \int [d\chi]\, e^{-S_\chi}    \e{pi}
\eqe
where
\eq
S_{\chi} = \beta\int d^d x\,\left(
\partial_i \chi^* \partial^i \chi + m^2(\beta) \chi^* \chi \right)\ ,
\eqe
and $d$ is the number of spatial dimensions. The field $\chi$ has
winding number one and $\chi^*$ has winding number minus one.
Eq.~\ee{pi} is the full multi-string partition function; the
single-string partition function is $Z = \ln {\bf Z}$.
The physical meaning of the thermal scalar has been a source of
confusion.  It has no apparent dynamical significance, but is useful in
determining the static properties of highly excited strings.

As an example, let us review Brandenberger and Vafa's use of the thermal
scalar to
calculate the density of states~\cite{bvafa}.  The log of the path
integral is
\eq
Z = - \sum_a \ln \lambda_a
\eqe
where the $\lambda_a$ are the eigenvalues of $-\nabla^2 + m^2(\beta)$.
When the sizes  of the spatial dimensions
are small compared to $m(\beta)^{-1}$, the splitting of the
$\lambda_a$ is large compared to the lowest eigenvalue
\eq
\lambda_1 = m^2(\beta)
\eqe
and this eigenvalue dominates the critical behavior,
\eq\e{smallpf}
Z_c(\beta)\ \approx\ - \ln \lambda_1\ \approx - \ln (\beta -
\beta_{\rm H})\ .
\eqe
This determines the density of states $n(M)$ for large mass:
\eq
n(M) = \frac{e^{\beta_{\rm H} M}}{M} 
\e{den1}
\eqe
where
\eq
Z(\beta) = \int_0^\infty
dM\,e^{-\beta M}n(M)\ .
\eqe
(Note that we are using
$M$ for the string mass and $m$ for the thermal scalar mass.)

When $d$ spatial dimensions are larger than $m(\beta)^{-1}$,
\eqa
Z(\beta)&\approx& - V \int \frac{d^dk}{(2\pi)^d}\, \ln \left(
k^2 + m^2(\beta) \right)
\nonumber\\
&\approx& V \int \frac{d^dk}{(2\pi)^d}\, \int_{0}^\infty
\frac{dM}{M}\, e^{-\beta M + \beta_{\rm H} M - 2\pi^2 \apm^2 k^2 
M/\beta_{\rm H}}
\nonumber\\
&=& \int_{0}^\infty {dM}\,e^{-\beta M} n(M)
\eqae
for
\eq
n(M) = V \frac{\beta_{\rm H}^{d/2}}{(4\pi^2\apm)^{d}}
\frac{ e^{\beta_{\rm H} M}}{M^{1 + d/2}}
 \ . \e{den2}
\eqe
We are ignoring divergences at $M \to 0$, which are ultraviolet from
the point of view of the effective field theory, but which relate to
the uninteresting light strings.

The thermal scalar also makes precise the random walk picture of the highly
excited string: in a first-quantized description, the $\chi$ path
integral is just the sum over random walks.
Consider for example the number of string states passing through
the origin and a second point $x$.  This is given by the thermal scalar path
integral as
\eq
\langle\, \chi^*\chi(x)\, \chi^*\chi(0)\, \rangle\ \sim\ e^{-2 |x|
m(\beta)} .
\e{corr}
\eqe
In the random walk picture, a string of energy $M$ is described by a gaussian 
whose width is proportional to $M^{1/2}$.  Averaging over the thermal ensemble
(only the exponential in the density of states is relevant) then
gives
\eq
\int_0^\infty dM\, e^{-x^2 C/ M} e^{-(\beta - \beta_{\rm
H} )M}\ \sim\ e^{-2|x|
 \sqrt{ C (\beta-\beta_{\rm H} ) } } .
\eqe
Indeed this has the same $x$ and $\beta$ dependence as the path integral
result~\ee{corr}, and determines $C = \beta_{\rm H} /2\pi^2\apm^2$.
The size of the random walk is then $l^2 = M/2C$.  Since also $l
\approx m(\beta)^{-1}$, the mass depends on the temperature
as $M \propto m(\beta)^{-2} \propto (\beta-\beta_{\rm H})^{-1}$.

The random walk picture also provides a simple explanation 
for the prefactors in the density of
states~\ee{den1} and~\ee{den2}.  The naive exponential count of the states
of a random walk overcounts by the length of the walk, since it is
irrelevant where along the loop the walk starts---hence the factor
$M^{-1}$ in the density~\ee{den1}.  In a large volume there is an
additional overcounting by the volume of the walk, $O(M^{d/2})$, because only
walks where the end coincides with the beginning are allowed.

\sect{Highly Excited Strings with Self Interaction}

We now wish to see how interactions modify the behavior of a typical
highly-excited string.  Since the string state is large compared to the string
scale, the most important interactions will be the long-ranged ones due to
exchange of gravitons and dilatons.  The statistical mechanics of
random walks with self-interactions is the subject of polymer
physics, and the scaling arguments we will make are similar to the
methods used in that subject~\cite{polymer}.  However, the case of a
polymer with a long-range attractive interaction has not previously
arisen.

Note that we are considering the
self-interaction of a single string, not the harder problem of the effect of
interactions on the full thermal ensemble at the Hagedorn transition. In
particular, there is no Jeans instability even though gravity will be important.
We will study the effect of interactions in a mean field approximation.  We
first determine the behavior a highly excited string in a fixed metric and
dilaton background, and then require that the background 
solve the field equations with
the typical string as source.

Consider a static dilaton $\Phi$ and static
string metric analytically continued to
imaginary time: $ds^2 = G_{\tau\tau} d\tau^2 + G_{ij} dx^i dx^j$.
The thermal scalar action in this background is
\eq
S_{\chi} = \beta \int d^dx\, \sqrt{G}e^{-2\Phi}
\left( G^{ij} \partial_i \chi^* \partial_j \chi + \frac{\beta^2 G_{\tau \tau} -
\beta_{\rm H}^2}{4\pi^2\apm^2} \chi^* \chi \right)\ , \e{tachact}
\eqe
The explicit factor of $G_{\tau\tau}$ is from the
proper length of the winding string.  The ${\tau\tau}$ component of the metric
also appears in $\sqrt{G}$ since this action can be obtained 
by dimensional reduction from a $d+1$ action.
The effective field theory of the low
energy degrees of freedom also includes the graviton-dilaton action
\eq
-\frac{\beta}{2\kappa^2} \int d^dx\ \sqrt{G}e^{-2\Phi} \left( R + 4 G^{ij}
\partial_i \Phi \partial_j \Phi \right)\ .
\eqe
Note that we are not interested in the full quantum field theory, which would
generate the full thermal ensemble.  Rather we want the single-string partition
function, corresponding to one random walk and so exactly one $\chi$ loop.
This can be written as a field theory by adding an index $a = 1, \ldots, n$ to
$\chi$ and taking the $n \to 0$ limit, but we will not use this formalism.

For weak fields,
the interactions between the dilaton  and thermal scalar in~\ee{tachact}
are suppressed by derivatives or $\beta-\beta_{\rm H}$. This is not true
for the metric, due to the explicit factor of $G_{\tau\tau}$.\footnote
{This is the string metric, so in the Einstein metric there are both
gravitational and dilaton forces.}  This cubic interaction is proportional to
the dimensionless string coupling $g$.  Other string interactions such as
the exchange of massive string excitations, or
a splitting-joining interaction of the long string, require the random walk to
intersect itself (or come within the string length), and so give rise to a
quartic interaction of the thermal scalar.  The exchange is order $g^2$ and
the splitting-joining of order $g$, but the quartic interaction is less
relevant than the cubic gravitational interaction and can be neglected.
Thus the dominant interaction is simply the gravitational
attraction of one part of the string on another.

We can make a simple estimate for when this interaction will be important.
The critical dimension for a cubic interaction is $d = 6$.  The coupling is
relevant for $d<6$, so we can anticipate that the effect of gravity
will be greater in lower dimensions. This is consistent with the fact that
the gravitational potential falls off more rapidly in higher dimensions.
A cubic coupling constant has units of
length$^{(d-6)/2}$ so the effective dimensionless coupling is
\eq
g m^{(d-6)/2}\ \sim\ g(\beta - \beta_{\rm H})^{(d-6)/4}
\ \sim\ gM^{(6-d)/4}\ ,
\eqe
temporarily omitting factors of $\apm$ to make the dependences clearer.  Thus
if we hold $N\sim M^2\apm$ large and fixed and increase $g$ from zero, the interaction
becomes important at
\eq
g_o \sim N^{(d-6)/8} \e{gone}
\eqe
for $d < 6$. Recall   that string scale black holes are formed when
$g \sim g_c \sim N^{-1/4}$. 
For $d=3$, we have $g_o < g_{c}$ so the interactions modify the free string
behavior in the weakly coupled regime.  For $d=4$, $g_o \sim g_{c}$ so the
interactions become important at the same scale where the localized strings
become black holes.  For $d = 5$, $g_{c} < g_o $ so the
interactions become important in the regime where the free strings are
metastable.

In an attractive potential, $G_{\tau\tau} < 1$ with $G_{\tau\tau} \to 1$
at infinity, one
expects the following effect. Locally, the effective temperature
$G_{\tau\tau}^{-1/2}
\beta^{-1}$ is increased, and the string can access more states than at
temperature
$\beta^{-1}$.  We would therefore expect the random walk to be concentrated in
the region of smallest
$G_{\tau\tau}$,
and the critical temperature $\beta_C^{-1}$ to be reduced relative
to
$\beta^{-1}_{\rm H}$.  This is the case, at least when the potential has a bound
state.  The operator
\eq
-(\nabla_\mu - 2\Phi_{,\mu}) \nabla^\mu +
\frac{\beta^2 (G_{\tau\tau} - 1)}{4\pi^2\apm^2} +
\frac{\beta^2 - \beta_{\rm H}^2}{4\pi^2\apm^2}
\e{op}
\eqe
then has lowest eigenvalue $\lambda_1$ less than the flat space value
$(\beta^2 -
\beta_{\rm H}^2) / 4\pi^2\apm^2 $.  As $\beta$ decreases from above, this
eigenvalue then vanishes at $\beta_C > \beta_{\rm H}$.  The
density of states then has the same form as in the small volume case above, but
with $\beta_C$ replacing $\beta_{\rm H}$,
\eq
n(M) = \frac{e^{\beta_{C} M}}{M} \ .
\e{den3}
\eqe
The bound state wavefunction gives the shape of the random walk.

The bound state picture gives a simple interpretation of the coupling $g_o$.
The condition that the operator~\ee{op} have a bound state   is roughly
\eq
\ell^2 V \roughly{>} 1\ , \e{boundcrit}
\eqe
where $\ell$ is the range of the potential and $V$ its
depth.  Taking the gravitational potential of a long string at its random walk
radius $\ell = N^{1/4}
\apm^{1/2}$, one has
\eq
\ell^2 V \sim G M \ell^{4-d} \sim g^2 N^{(6-d)/4}\ .
\eqe
The bound state criterion~\ee{boundcrit} is then $g \roughly{>} g_o$.

We now wish to require that the background satisfy the field equations with 
sources coming from the excited string. In the mean field approximation,
we average these sources over all excited strings with the same mass:
\eqa
R + 4 \nabla^2 \Phi - 4\nabla_\mu \Phi \nabla^\mu \Phi &=& 2\kappa^2
\langle J\rangle\ .
\nonumber\\ R_{\mu\nu} + 2\nabla_\mu \nabla_\nu \Phi &=& \kappa^2 
\left [e^{2\Phi}
\langle T_{\mu\nu} \rangle +  G_{\mu\nu} \langle J\rangle \right ]\ ,
\e{eom}
\eqae
where $J$ is the quantity in parentheses in the scalar 
action~\ee{tachact}. It is shown in the Appendix
that the sources are simply given by the classical expression evaluated at
the bound state wavefunction $\chi$, times an appropriate normalization.
The wavefunction satisfies 
\eq
\left\{ -(\nabla_\mu - 2\Phi_{,\mu}) \nabla^\mu +
\frac{\beta_{C}^2 (G_{\tau\tau} - 1)}{4\pi^2\apm^2} +
\frac{\beta_{C}^2 - \beta_{\rm H}^2}{4\pi^2\apm^2} \right\} \chi = 0\ .
\e{eigen}
\eqe

The low energy field equations \ee{eom} are
valid only when all derivatives are small compared to the string scale. 
Due to the explicit factors of $\apm$ in the eigenvalue equation \ee{eigen}
this requires the further approximation 
\eqa
\beta_{C}^2 - \beta_{\rm H}^2 &\ll& 1 \nonumber\\
h_{\tau\tau}\ \equiv\  G_{\tau\tau}- 1  &\ll& 1\ . \e{red}
\eqae
Thus we can linearize the equations for the background. In the usual
Lorentz gauge, $R_{\mu\nu} = -{1\over 2} \partial^2 h_{\mu\nu}$.
To leading order,
$\langle J \rangle =0$, so the dilaton equation becomes $R + 4\partial^2 \Phi
=0$ with solution $ \Phi = {h_\mu}^\mu/8$. 
The $R_{\tau\tau}$ equation reduces to Newton's law
\eq
\partial^i \partial_i h_{\tau\tau} = 2\kappa^2 M \chi^* \chi
\eqe
where we have imposed the normalization $\int d^d x \chi^*\chi =1$.
Solving this, the eigenvalue equation~\ee{eigen}
becomes
\eq
-\partial^i \partial_i \chi(x)
- \frac{\beta_{\rm H}^2 \kappa^2 M }{2\pi^2 (d-2)\omega_{d-1} \apm^2} \chi(x)
\int d^dx'\,\frac{\chi^*\chi(x')}{x_{\mbox{\tiny $>$}}^{d-2}}\ =\
\frac{ \beta_{\rm H}( \beta_{\rm H} - \beta_C ) }{2\pi^2\apm^2} \chi(x)
\e{nlschrod}
\eqe
where $x_{\mbox{\tiny $>$}}$ is the greater of $|x|$ and $|x'|$ and
$\omega_{d-1}$ is the volume of the unit $S_{d-1}$.

Eq.~\ee{nlschrod} is a nonlinear Schr\"odinger equation with attractive Coulomb
self-interaction.  The essential physics can be obtained by a scaling argument.
Define a dimensionless string coupling
\eq
g^2\ =\ \frac{\beta_{\rm H}^2 \kappa^2 }{2\pi^2 (d-2)\omega_{d-1} \apm^2} 
\eqe
and rescaled variables\footnote
{This preserves the normalization $\int d^dx\,\chi^*\chi =
\int d^dy\,\psi^*\psi = 1$.}
\eq
x = (g^2M)^{1/(d-4)} y, \qquad \chi(x) = (g^2M)^{d/(8-2d)}
\psi(y)
\eqe
(setting aside temporarily the case $d=4$).
The eigenvalue equation~\ee{nlschrod} becomes
\eq
-\partial_{y^i} \partial_{y^i} \psi(y)
-  \psi(y)
\int d^dy'\,\frac{\psi^*\psi(y')}{y^{d-2}_{\mbox{\tiny $>$}}}\ =\
\zeta \psi(y)\
\e{scale}
\eqe
where
\eq
\beta_{\rm H} - \beta_C = \zeta \frac{2 \pi^2 \apm^2 (g^2 M)^{2/(4-d)}}
{\beta_{\rm H}}\ .
\eqe

We are interested in the lowest bound state solution to this equation.
Formally this can be found by minimizing
\eq
I = \int d^dy\, \partial_{y^i} \psi^* (y)\partial_{y^i} \psi(y)
- \frac{1}{2} \int d^dy\,
\int d^dy'\,\frac{\psi^* \psi(y) \psi^*\psi(y')}{y^{d-2}_{\mbox{\tiny $>$}}}
\eqe
subject to $\int d^dy\,\psi^*\psi = 1$.
However, we need to be sure that this functional is bounded from below. 
If we rescale
$\psi(y) \to \rho^{d/2} \psi(\rho y)$, the first (positive) term scales as
$\rho^{2}$ and the second (negative) term scales as $\rho^{d-2}$.  The negative
term becomes arbitrarily large as $\rho \to \infty$ (for $d\geq 3$).  For $d
= 3$ the positive term grows faster in this limit and so the variational
principle predicts a lowest bound state.  For $d \geq 5$, $I$ can
be arbitrarily negative and there is no state of lowest eigenvalue.  For $d=4$ one can
perform the scaling in the original variables~\ee{nlschrod}.  The two terms
both scale as $\rho^{2}$ so the coupling does not scale out.  For small
coupling the kinetic term dominates and $I$ is positive.  Past a critical
coupling the potential dominates and $I$ can decrease without bound.

Let us first consider the case $d=3$.
Since all constants have been
scaled out of eq.~\ee{scale}, we expect the lowest eigenvalue to be $\zeta_0
\sim  O(-1)$ and
the size of the bound state to be $O(1)$ in the $y$ variable.  In terms of the
original variables, this gives
\eqa
\beta_C - \beta_{\rm H} &\sim&  \frac{ \apm^2 g^4 M^2}
{\beta_{\rm H}}\nonumber\\
\ell &\sim& g^{-2} M^{-1}\ . \e{size}
\eqae
We can also express this in terms of the excitation level of the string.
Because the redshift~\ee{red} is small, the mass--level relation is
approximately as in the free case,
\eq\e{massfree}
M^2 = \frac{4}{\apm} N\ .
\eqe
Thus the size~\ee{size} of the string state is of order
\eq\e{size3d}
\ell \sim { \apm^{1/2}\over g^2 N^{1/2}} \ .
\eqe
This is one of our main results. It is nonperturbative in
the coupling $g$, and is valid for $g_o < g < g_c$.
Since $g_o \sim  N^{-3/8}$ and $g_c \sim N^{-1/4}$, it shows that the
size of a typical excited  string in three spatial dimensions smoothly
interpolates from the random walk size to the string scale as one
slowly increases the coupling. For
$g< g_o$, the interactions are negligible and the string remains at its
random walk size. The result \ee{size3d} is not applicable since
there is no bound state. For
$g>g_c$ the string forms a black hole. If one increases the coupling further,
the black hole size will be fixed to be $N^{1/4}$ in Planck units,
but grow like $g N^{1/4}$
in string units.

For $d=4$ and $d=5$, once the string coupling reaches $g_o$ the
estimate~\ee{boundcrit} indicates that  bound states form, but we have seen
that the system becomes unstable: there are states of arbitrarily negative
energy.  We interpret this as saying that once the interaction becomes
important the long string collapses all the way to a black hole.  For $d=6$ the
interaction is marginal and for $d>6$ it is irrelevant, but this does not mean
that it can be neglected.  These terms refer to the scaling if we hold $g$ fixed
and increase the length scale---that is, $M$.  However, we are holding $M$ fixed
and increasing $g$.  In this case, one always reaches the coupling $g_o$ where
a bound state can form.  Again, it is unstable and should collapse.  For $d >
6$, $g_o \gg 1$ and the theory is out of the range of validity of the original
theory. One can still discuss the evolution of the string by passing to a
weakly coupled dual theory. The original fundamental string becomes a 
solitonic  string with  tension that increases as the dual coupling $\tilde g
=1/g$ is decreased. If this state does not decay, one might expect it to
form a black hole when $\tilde g \sim 1/g_o$. However, the dual theory
has much lighter degrees of freedom---long dual strings, for example---with much
higher entropy at given mass. If the solitonic string rapidly decays into these
dual strings, then a  black hole will not form. This would imply that
most excited states of strings in higher dimensions never form a black hole
for any value 
of the string coupling. However,
the decay of the solitonic string to the dual strings might be quite
slow, because it is locally a BPS state: small loops must break off and
contract for it to decay.

\sect{Appendix: Calculation of $\langle T_{\mu\nu}\rangle$}

In this appendix, we compute the mean value of the stress  energy tensor
among all string states with mass $M$.
First we represent this  tensor as a functional
derivative of the string Hamiltonian:
\eq
T^{\mu\nu} = -\frac{2}{\sqrt{G}} \frac{\delta H}{\delta G_{\mu\nu}}\ .
\eqe
Its expectation value in a typical state of mass $M$ is then
\eqa
\langle\, T^{\mu\nu}\, \rangle &=& \frac{ {\rm Tr}\left\{ T^{\mu\nu}
\delta(H-M) \right\} }{ {\rm Tr} \left\{
\delta(H-M) \right\} } \nonumber\\
&=& \frac{ 2 }{ \sqrt{G}{\rm Tr}\left\{ \delta(H-M)\right\} }
\frac{ \delta}{ \delta G_{\mu\nu} }{\rm Tr} \left\{ \theta(M - H) \right\}
\eqae
Evaluating the traces using the density of states~\ee{den3} gives
\eqa
\langle\, T^{\mu\nu}\, \rangle &\approx& \frac{2}{\sqrt{G}}
e^{-M\beta_C}
\frac{\delta}{\delta G_{\mu\nu}} \left( \beta_C^{-1}
e^{M\beta_C}\right)  \nonumber\\
&\approx& \frac{M }{\beta_C^2 \sqrt{G}}\frac{\delta \beta_C^2}{\delta
G_{\mu\nu}}\ . \e{tmunu}
\eqae
where these expressions are valid in the limit of large $M$.
The critical temperature $\beta_C$ was defined by
$\lambda_1(G_{\mu\nu},\beta_C) = 0$, so
\eq
\frac{\delta \beta_C^2}{\delta G_{\mu\nu}} =
- \frac{ \delta \lambda_1/\delta  G_{\mu\nu}}{ \delta \lambda_1 /\delta
\beta^2 |_{\beta = \beta_C}}\ .
\eqe
The derivatives of the eigenvalues are given by first-order perturbation
theory,
\eqa
\frac{e^{2\Phi}}{\sqrt{G}} \frac{\delta \lambda_1}{\delta G_{\tau\tau}} &=&
 \frac{\beta_C^2 \chi^* \chi }{ 4\pi^2\apm^2 } + \frac{1}{2} G^{\tau\tau} J
\nonumber\\
\frac{e^{2\Phi}}{\sqrt{G}} \frac{\delta \lambda_1}{\delta G_{ij}} &=&
-\nabla^{(i} \chi^* \nabla^{j)}\chi + \half G^{ij} J  
\nonumber\\
\frac{\delta \lambda_1}{\delta \beta^2} &=& \int d^dx\,\sqrt{G} G_{\tau\tau}
e^{-2\Phi} \frac{\chi^* \chi }{ 4\pi^2\apm^2}\ , \e{source}
\eqae
where $\chi$ is a solution to \ee{eigen}, and $J$ 
is the quantity in
parentheses in the action $S_\chi$ \ee{tachact}
evaluated on the bound state wave function. The resulting stress energy
tensor is simply the  variation of $S_\chi$ with respect to the metric,
evaluated on the bound state wave function.     

The stress energy tensor~\ee{tmunu} satisfies two important consistency 
checks. Since it does not include the stress energy of the dilaton
field, it is not conserved by itself. Instead it satisfies
\eq
\nabla_\mu \langle T^{\mu\nu} \rangle =2e^{-2\Phi}\langle J\rangle \nabla^\nu
\Phi\ .
\eqe
This is required for the consistency of the field equations and Bianchi
identities, and is a necessary check because the action $S_\chi$ is not
manifestly invariant under time reparameterizations.
It is also correctly
normalized in the following sense. In  a static spacetime, the total
energy associated with the matter is
\eq
M_{matter} = \int_\Sigma T_{\mu\nu} \xi^\mu n^\nu d\Sigma
\eqe
where $\xi^\mu$ is the timelike Killing vector, and
the integral is over a static surface $\Sigma$ with unit normal 
$n^\nu$ and proper volume $d\Sigma$. Using the above expression for 
$\langle T_{\mu\nu} \rangle$, we find\footnote{Recall that $\sqrt {G}$
includes the $\tau\tau$ component of the metric.}
\eq
M_{matter} = -\int \langle {T_\tau}^\tau\rangle \sqrt {G} d^d x = M
\eqe

The total ADM mass of a static spacetime
can be expressed similarly in terms of an integral of the Ricci tensor 
rather than the stress energy tensor. Assuming $D$ spacetime dimensions,
and using the Einstein metric, one has 
\eq
M_{ADM} = {D-2\over (D-3) \kappa^2}\int_\Sigma
\tilde R_{\mu\nu} \xi^\mu \tilde n^\nu d\Sigma
\eqe
This differs from $M_{matter}$ since it also includes the gravitational
binding energy.
Rewriting this expression in terms of the string metric yields
\eqa&&\hspace{-20pt}
M_{ADM} = {D-2\over (D-3) \kappa^2}\int_\Sigma
e^{-2\Phi}\biggl[ R_{\mu\nu} + 2\nabla_\mu \nabla_\nu \Phi
\\
&&\qquad\qquad\qquad +\, {2\over D-2} 
G_{\mu\nu} ( \nabla^2 \Phi -2\nabla_\mu \Phi \nabla^\mu\Phi)\biggr ]
\xi^\mu  n^\nu d\Sigma\ .  \nonumber
\eqae
Using the equations of motion \ee{eom} this becomes
\eq
M_{ADM} = {D-2\over D-3} \int \left[\langle T_{\mu\nu}\rangle -G_{\mu\nu} 
{ \langle{T_\alpha}^\alpha\rangle\over D-2} \right] \xi^\mu  n^\nu d\Sigma
\eqe
For weak
fields, one recovers $M_{ADM} = M$. In general, these two masses will not be 
equal, but even when a black hole is about to form, 
they will differ only by a factor
of order unity. 

In terms of the correspondence principle, if one starts at zero coupling
with a (compact) string state of mass $M$,  it will form a black hole at larger
coupling with mass
$M_{ADM}$. Since these two masses
differ only by a numerical factor, 
the black hole entropy is reproduced (up to a similar factor) even
if the black hole mass is equated with the string mass at  zero coupling.

\subsection*{Acknowledgments}
It is a pleasure to thank J. Cardy and A. Strominger for discussions.
This work was supported in part by NSF grants PHY94-07194, PHY95-07065,
and PHY97-22022.

\end{document}